\documentclass[conference]{IEEEtran}
\IEEEoverridecommandlockouts

\usepackage{cite}

\ifCLASSINFOpdf
  \usepackage[pdftex]{graphicx}
  \DeclareGraphicsExtensions{.pdf,.jpeg,.jpg,.png}
  \usepackage[protrusion=true,expansion=true]{microtype}  

\else
  \usepackage[dvips]{graphicx}
  \DeclareGraphicsExtensions{.eps,.jpeg,.jpg,.png}
  \usepackage{microtype}
\fi
\usepackage{amsmath}
\usepackage{amssymb}
\usepackage{algorithm}
\usepackage[noend]{algpseudocode}

\usepackage{array}

\usepackage{todonotes}
\usepackage{multirow}
\usepackage{gensymb} 

\newcommand{\integral}[4]{
\int_{#3}^{#4}  #1 \, \mathrm{d}#2 }

\newcommand{\e}[1]{ \mathrm{e}^{#1}}

\renewcommand{\log}[1]{ \mathrm{log}\left( #1 \right)}

\newcommand{\Ton}[0]{T_{\mathrm{on}}}
\newcommand{\Toff}[0]{T_{\mathrm{off}}}
\newcommand{\Tmax}[0]{T_{\mathrm{max}}}
\newcommand{\Tmin}[0]{T_{\mathrm{min}}}
\newcommand{\Tlow}[0]{T_{\mathrm{low}}}
\newcommand{\Thigh}[0]{T_{\mathrm{high}}}

\begin{document}

%
\title{\Large \textbf{Low-complexity decentralized algorithm for aggregate load control \\of thermostatic loads}}

\author{\IEEEauthorblockN{\Large Simon H. Tindemans}
\IEEEauthorblockA{Member, IEEE\\
Department of Electrical Sustainable Energy\\
Delft University of Technology\\
Delft, The Netherlands\\
s.h.tindemans@tudelft.nl}
\and
\IEEEauthorblockN{\Large Goran Strbac}
\IEEEauthorblockA{Member, IEEE\\
Department of Electrical and Electronic Engineering\\
Imperial College London\\
London, United Kingdom\\
g.strbac@imperial.ac.uk}
\thanks{This research was supported by EDF Energy R\&D UK Centre as part of the STAMINA project.}}


%


\IEEEoverridecommandlockouts

\IEEEpubid{\begin{minipage}{\textwidth}\copyright 2020 IEEE.  Personal use of this material is permitted.  Permission from IEEE must be obtained for all other uses, in any current or future media, including reprinting/republishing this material for advertising or promotional purposes, creating new collective works, for resale or redistribution to servers or lists, or reuse of any copyrighted component of this work in other works.
\end{minipage}}

\maketitle



\IEEEpubidadjcol

\begin{abstract}
Thermostatically controlled loads such as refrigerators are exceptionally suitable as a flexible demand resource. This paper derives a decentralised load control algorithm for refrigerators. It is adapted from an existing continuous time control approach, with the aim to achieve low computational complexity and an ability to handle discrete time steps of variable length -- desirable features for embedding in appliances and high-throughput simulations.  Simulation results of large populations of heterogeneous appliances illustrate the accurate aggregate control of power consumption and high computational efficiency. Tracking accuracy is quantified as a function of population size and time step size, and correlations in the tracking error are investigated. The controller is shown to be robust to errors in model specification and to sudden perturbations in the form of random refrigerator door openings.  
\end{abstract}

\renewcommand\IEEEkeywordsname{Keywords}


%
\IEEEpeerreviewmaketitle

\section{Introduction}

The physical characteristics of refrigerators and other thermostatically controlled loads  (TCLs) make them exceptionally suitable as a low-cost provider of flexibility to the grid: their power consumption can be shifted by tens of minutes without noticeable effects on cooling performance. This flexibility can then be used for the provision of response and reserve services, to reduce extreme load levels and to alleviate ramping constraints \cite{Callaway2011}. 

The fundamental challenge in deploying TCLs as a flexibility resource stems from the fact that most such loads only have high-power and low-power states (although exceptions exist, and lead to distinct challenges \cite{Vorwerk2020}). Although the concept of frequency response from such \texttt{on}/\texttt{off} TCLs has been around for decades \cite{Schweppe1979}, it has gained popularity in recent years, since Short \emph{et al.} illustrated how a simple adaptation of a hysteresis controller could be used to contribute to frequency stability \cite{Short2007}. The simple control mechanism is effective, but may cause synchronisation of appliances that are modulated by a common frequency signal \cite{Angeli2012}. Ad hoc randomisation strategies have been proposed to prevent this problem \cite{Hirst2000}, but recently Webborn \cite{Webborn2019} has argued that the inherent heterogeneity of appliances is sufficient to dampen this effect in practical applications. 

A second development has been a desire to develop controllers than can not only stabilise the grid, but accurately track regulation signals. This can be achieved in a (conceptually) straightforward manner using a real-time communication infrastructure for direct measurement and control of appliances (e.g. \cite{Hao2015a}). However, considerations of privacy and fault-tolerance mean that it is generally preferred to consider systems that are only reliant on global or infrequent measurements. This can be achieved by incorporating models of device-level temperature dynamics \cite{Chong1979, Ihara1981} and considering aggregate population dynamics, for example in the form of Markov transition models (`temperature bin models') \cite{Mathieu2013a} that are used to improve aggregate state estimation. 

\IEEEpubidadjcol
\IEEEpubidadjcol

Moreover, since many TCLs are deployed in large numbers, control of their aggregate power is a natural application for randomised control schemes. An early such scheme \cite{Angeli2012} has the drawbacks that it controls only asymptotic power consumption and cannot safeguard the temperature constraints of individual appliances. More recent papers employing randomised control \cite{Tindemans2015, Vrettos2016b, Totu2017, Chertkov2017, Busic2019} employ an explicit representation of temperatures to prevent unwanted temperature excursions. Within these, \cite{Vrettos2016b, Totu2017} focus on realism, incorporating details such as as limits on compressor cycling and non-constant power demand. On the other hand, \cite{Busic2019} focuses on the development of a more general continuous-time control framework.

This paper extends the decentralised randomised control strategy that was introduced in \cite{Tindemans2015, Trovato2015}. That strategy has the desirable feature that it requires only one-way broadcast information, yet achieves tracking of a reference signal that is immediate and exact in expectation (i.e. exact for large numbers) without violating temperature limits in individual devices. However, its continuous time formulation in integral form was not conducive to implementation on embedded controllers or for rapid simulation of many devices at once. The importance of such performance considerations for practical implementations can be seen in \cite{Iacovella2016, PoncedeLeonBarido2017}. Against this background, the contributions of the present paper are as follows:
\begin{itemize}
\item The distribution-referred approach to aggregate load control is concisely described in Section~\ref{sec:preliminaries} and the associated equations are restated in natural coordinates in Section~\ref{sec:distribution}. 
\item A discrete-time controller is derived that is designed to track a piecewise-constant reference signal. It consists of two complementary parts: \emph{distribution control} (Section~\ref{sec:distribution}) and \emph{device switching}  (Section~\ref{sec:switching}). The algorithm is particularly suitable for implementation on devices with computational constraints. Specifically, it avoids numerical integration and uses greedy time steps of variable size, so that real-time performance requirements are relaxed. 
\item An explicit control algorithm is stated (Section~\ref{sec:algorithm}) and an open source Python implementation is made available.
\item Simulations confirm the ability of the proposed algorithm to track a (potentially discontinuous) reference signal with a heterogeneous population of TCLs. The tracking error reduces with fleet size.  
\item The tracking error is quantified as a function of time step size, and correlations in the tracking error are investigated.
\item The controller is demonstrated to be robust to errors in model specification and to sudden state perturbations in the form of random refrigerator door openings. 
\end{itemize}
This paper is an extended version of the conference paper \cite{Tindemans2019}, which first introduced the discrete-time formulation of the controller and demonstrated its tracking performance for large fleets (contributions 1-4). Compared to the earlier paper, this paper includes an extended discussion of the probabilistic formalism underlying the controller, a functionally equivalent but more compact statement of the control algorithm, and additional analysis of the controller performance (contributions 5-6). 

\section{Preliminaries} \label{sec:preliminaries}

\subsection{Appliance model}

Throughout this paper, we consider a first order TCL model \cite{Chong1979, Ihara1981}, expressed by the following differential equation \cite{Angeli2012} for the temperature $T^a$ of the compartment of appliance $a$:
\begin{equation} \label{eq:ode-physics}
\frac{\mathrm{d} T^a(t)}{\mathrm{d}t} = - \alpha^a \left[ T^a(t) - T^a_{\mathrm{off}} + c^a(t) \cdot \left(  T^a_{\mathrm{off}} - T^a_{\mathrm{on}} \right)  \right]
\end{equation}
Here, $c^a(t) \in \{0,1\}\equiv\{\texttt{off}, \texttt{on} \}$ is the state of the compressor, and $T^a_{\mathrm{off}}$ and $T^a_{\mathrm{on}}$ are the asymptotic temperatures in the \texttt{off} and \texttt{on} states, respectively. For ease of exposition, we shall refer to refrigerating applicances throughout, although the same model and control strategy can be used for other TCLs, e.g. space heaters. For the refrigeration case, the model parameters are related to physical parameters as follows \cite{Vrettos2016b}: $\alpha^a=1/(R^a C^a)$, where $C^a$ is the thermal capacitance and $R^a$ is the thermal resistance. $\Toff^a$ is the ambient temperature and $\Ton^a = \Toff^a - \eta^a R^a P^a_{\textrm{on}}$, with $\eta^a$ the coefficient of performance and $P^a_{\textrm{on}}$ the work performed by the compressor. The power consumption $P^a(t)$ of the appliance is assumed to be dominated by the compressor power consumption $P^a_{\textrm{on}}$, so that
\begin{equation}
P^a(t) = P^a_{\textrm{on}} c^a(t).
\end{equation}

In the steady state (no control actions), the appliance is subject to a hysteresis controller that switches to the \texttt{on} state whenever the upper temperature bound $\Tmax^a$ is reached, and to the \texttt{off} state when $\Tmin^a$ is reached. This results in a periodic cycling of the power consumption, with an average power level $P^a_0$. 

We can transition to a probabilistic framing of the control approach in the following manner. Let us consider that the appliance $a$ has a known model $\mathcal{M}^a =( \alpha^a, P^a_{\textrm{on}}, \Toff^a, \Ton^a, \Tmin^a, \Tmax^a )$. A \emph{hypothetical} large population of independent devices with identical models $\mathcal{M}^a$ would have internal states $(T^a(t), s^a(t))$ that are distributed according to a steady state distribution $\mathcal{D}^a_0(\mathcal{M}^a)$. Now consider a \emph{single} appliance $a$, with an unknown (to the external observer) internal state $\mathcal{S}^a = (T^a(t), c^a(t) )$, i.e. an unknown initial condition. We can then say that its state is a random variable that is drawn from the steady state distribution, $\mathcal{S}^a \sim \mathcal{D}^a_0(\mathcal{M}^a)$. Moreover, if appliances are independently operated, this state is independent of the states of other appliances (regardless of whether their models are different). Then, the power consumption of device $a$ is in effect a random process, with the (constant) expectation
\begin{equation} \label{eq:expectationsingle}
E_{\mathcal{S}^a(t)}[P^a(t)] = P^a_0,
\end{equation}
where \cite[Eq.~(33)]{Tindemans2015}
\begin{equation}
P^a_0 = \left[ \frac{  \log{\frac{\Tmax^a-\Ton^a}{\Tmin^a - \Ton^a}}}{\log{ \frac{(\Tmax^a-\Ton^a)
   (\Tmin^a-\Toff^a)}{(\Tmin^a-\Ton^a)(\Tmax^a-\Toff^a) }}}\right] P^a_{\textrm{on}}. 
\end{equation}
The expectation \eqref{eq:expectationsingle} is therefore taken over the internal states that an appliance might have at time $t$, given its model $\mathcal{M}^a$. Because no external control is applied and the initial state of the appliance was unknown, the result does not depend on time.

\subsection{Aggregate power modulation}

The objective of the TCL demand response controller is to control the aggregate power consumption 
\begin{equation} \label{eq:sumpower}
P_{\Pi}(t) = \sum_{a \in \mathcal{A}} P_{\Pi}^a(t)
\end{equation}
of a collection of appliances $\mathcal{A}$. In this paper we consider the control approach introduced in \cite{Tindemans2015}, which modulates the power consumption using a broadcast signal $\Pi(t)$. This \emph{reference signal} is available to all devices, and indicates the desired power consumption relative to the nominal power consumption of the device. Each device individually adapts its power consumption $P^a_{\Pi}(t)$ in order to satisfy 
\begin{equation} \label{eq:expectationpi}
E_{\mathcal{S}^a(t)}[P^a_{\Pi}(t)] = \Pi(t) P^a_0.
\end{equation}
By design, a signal $\Pi(t)=1$ results in the steady state power consumption \eqref{eq:expectationsingle}, and changes in $\Pi(t)$ are immediately reflected in the expected power consumption. Moreover, the controller maintains independence between appliances: the random states $\mathcal{S}^a(t)$ and $\mathcal{S}^b(t)$ are independent of each other, when conditioned on the control signal $\Pi(\cdot)$. It thus follows from the Lyapunov condition (Lyapunov central limit theorem) that the aggregate power consumption \eqref{eq:sumpower} is given by
\begin{equation} \label{eq:sumconvergence}
P_{\Pi}(t) = \Pi(t) \sum_{a \in \mathcal{A}} P^a_0 + O(\sqrt{|\mathcal{A}|}),
\end{equation}
where the last term is a random process that decreases in relative importance to the first term as the set of appliances increases. We note that this is the case even for heterogeneous appliances. 

The ability to closely track a reference signal was first demonstrated in \cite{Tindemans2015}. In \cite{Trovato2015}, the control signal was generated using a mixture of off-line scheduling and real-time control, and in \cite{Tindemans2015a} various frequency-sensitive controllers to locally compute $\Pi(t)$ (e.g. a simple droop controller) were implemented.

\subsection{Distribution-referred control} \label{sec:drcontrol}

The controller introduced in \cite{Tindemans2015, Trovato2015} can be thought of as  \emph{distribution-referred}, because a device first derives a generic control model for devices of \emph{its type}, by considering its thermal model $\mathcal{M}^a$ without reference to its actual state $(T^a(t), c^a(t))$. In accordance with the probabilistic frame defined in the previous section, each device considers a probability distribution of temperatures for appliances with its model $\mathcal{M}^a$ and an unknown state $\mathcal{S}^a$. One convenient approach is to parameterise this temperature distribution as a family $f_z(T;\mathcal{M}^a)$ that varies continuously in the parameter $z$, containing as a special case the steady state temperature distribution \cite[Eqs.~(31)-(32)]{Tindemans2015}
\begin{equation}
f_0(T; \mathcal{M}^a) = \frac{k^a}{(\Toff^a - T)(T-\Ton^a)}
\end{equation}
with 
\begin{equation} \label{eq:c-constant}
k^a = \frac{\Toff^a-\Ton^a}{\log{ \frac{(\Tmax^a-\Ton^a)
   (\Tmin^a-\Toff^a)}{(\Tmin^a-\Ton^a)(\Tmax^a-\Toff^a) }}}.
\end{equation}

At a high level, the controller for each appliance consists of two parts, which are evaluated in order:
\begin{enumerate}
\item \textbf{Distribution control.} Control the evolution of the distribution parameter $z(t)$ in such a way that the power consumption tracks the reference signal $\Pi(t)$ according to \eqref{eq:expectationpi}. Determine the collective device switching actions required to keep device temperatures aligned with $f_{z(t)}(T)$, and identify the temperature limits $\Tmin(t) \ge \Tmin$ and $\Tmax(t) \le \Tmax$.
\item \textbf{Device switching.} Based on the \emph{actual} appliance state $(T^a(t), c^a(t))$, compute stochastic control actions, in the form of \texttt{on}/\texttt{off} switching. Switching events can be initiated in three distinct ways:
\begin{itemize}
\item Forced switching (deterministic) when temperature limits $\Tmin(t)$ or $\Tmax(t)$ are exceeded.
\item A continuous-time switching process (stochastic) at intermediate temperatures in order to shape the temperature distribution.
\item Instantaneous switching (stochastic) on discrete changes of power setpoints, or when the controller switches between energy-provision and energy-absorption modes \cite{Trovato2015} (see also Section~\ref{sec:controllermodes}).
\end{itemize}
\end{enumerate}
The distribution control and device switching phases for the discrete time control strategy are addressed in Sections \ref{sec:distribution} and \ref{sec:switching}, respectively. We will henceforth drop the appliance superscript $a$, because the control steps are executed locally within each appliance (or independently for each appliance in a simulation).  Note that this implies a single model $\mathcal{M}$ is used in the derivations, but the results remain valid for portfolios of heterogeneous devices, each with their own model.

\subsection{Time-discretisation procedure}

The derivation of the controller in \cite{Tindemans2015} was performed in continuous time and temperature coordinates. However, implementation in device controllers is greatly simplified when a discrete time formulation can be given, which is the main contribution of this paper. To move from a continuous-time to a discrete-time formulation, we partition the timeline by the ordered sequence of times $\{t_i\}$, indexed by the integer $i$, at which the controller is invoked. These define time intervals $(t_{i-1}, t_i]$ with durations $\Delta t_i =t_i - t_{i-1}$. Note that the duration $\Delta t_i$ refers to the interval \emph{prior} to $t_i$, and the intervals may have variable size. The reference signal $\Pi(t)$ is assumed to be piecewise constant, defined by 
\begin{equation}
\Pi(t) = \Pi_i, \quad \text{for } t\in (t_{i-1}, t_i].
\end{equation}
The controller thus receives at $t_i$ a new reference power level $\Pi_{i+1}$ that must be applied for the upcoming interval $(t_i,t_{i+1}]$. Note that $t_{i+1}$ is not necessarily known at time $t_i$. 

Although the discontinuous changes of reference power will trigger switching events at $t_i$, the other switching events may occur at any time $t$. In the discretised approximation of the continuous time controller, they will be synchronised with the control execution times $t_i$ as follows. It is assumed that switching is immediate (at $t_i$).
\begin{itemize}
\item A violation of the temperature limits will trigger corrective switching as soon as it is detected.
\item Switching between energy-provision and energy-absorption modes also results in a discontinuous change in the desired local heating/cooling rate. This is achieved by on/off switching, immediately when a change in regime is detected.
\item Continuous time stochastic switching is implemented by approximating the integrated switching rate (i.e. the switching probability) over $\Delta t_i$ using the trapezoidal method, and executing any switching events at $t_i$ (the end of the interval). 
\end{itemize}

The algorithm is thus implemented in a `backward' fashion, meaning that at time $t_i$, the algorithm implements switching actions resulting from reference changes at $t_i$, and those accumulated over the preceding interval $(t_{i-1}, t_i]$. The advantage of this approach is that the interval $\Delta t_i$ can be chosen opportunistically: the controller does not need to know in advance when the next time step will take place. This is convenient, for example when computational limitations cause a delay in intended invocation time, or when the time step adapts to sudden changes in grid frequency. It should be pointed out that this `backward' integration does not delay the response to changes in reference power, which is implemented immediately at $t_i$. 

\section{Controller: manipulation of distribution} \label{sec:distribution}

This section focuses on the first part of the controller. It computes the desired evolution of the probability distribution of temperatures of fridges with model $\mathcal{M}$, when tracking a piecewise constant reference $\Pi(t)=\Pi_i$, for $t \in (t_{i-1},t_i]$. The derivation is initially performed in continuous time. The results are subsequently expressed in natural coordinates and restated in a form that is suitable for discrete-time evaluation. 

\subsection{Aggregate physics}

The average temperature of a TCL population is affected by the desired power consumption $\Pi(t)$ according to \cite[Eq.~(26)]{Tindemans2015}. With the convention that $t_{-1}=-\infty$ and $\Pi_0 = 1$ (assuming an initial steady state), it follows that
\begin{align}
\overline{T}(t_i) 
&= \Toff - \alpha (\Toff - \overline{T}_0) \sum_{j=0}^{i} \Pi_j \integral{ \e{-\alpha (t_i-t')}}{t'}{t_{j-1}}{t_j}, \label{eq:avgtempdefinition}
\end{align}
where the steady state average temperature $\overline{T}_0$ is computed using \cite[Eqs.~(23) and (32)-(33)]{Tindemans2015} as
\begin{equation}
\overline{T}_0 = \Toff - k \times \log{\frac{\Tmax-\Ton}{\Tmin - \Ton}} \label{eq:steadyStateTemperature}
\end{equation}
with $k$ defined in \eqref{eq:c-constant}. We define the dimensionless variable
\begin{equation}
z(t) =\frac{\overline{T}_0 - \overline{T}(t)}{\Toff - \overline{T}_0}, \label{eq:zdefinition}
\end{equation}
to parameterise the distributions $f_{z(t)}(T)$, and simplify the notation in what follows. $z$ is a representation of the cooling energy stored in the aggregate device population, relative to the uncontrolled state ($z=0$).\footnote{Note that $z$ is related to the variable $\sigma$ used in \cite{Trovato2015} as $z=\sigma-1$.} 

\subsection{Controller modes} \label{sec:controllermodes}

The algorithm in \cite{Tindemans2015} implicitly generates the family of temperature distributions $f_{z(t)}(T)$ by the net heating rate $v(T,t)$, which is determined by averaging the heating/cooling rate over devices in the \texttt{off} (heating) state and \texttt{on} (cooling) state at time $t$ and temperature $T$. The net heating rate and $f_{z(t)}$ are coupled through the continuity equation
\begin{equation}
\frac{\partial }{\partial t}f_{z(t)}(T) =  -\frac{\partial}{\partial T}\left[ v(T,t) f_{z(t)}(T) \right].  \label{eq:PDEsymm}
\end{equation}

The heating rate can be chosen freely between the lower and upper limits that are defined by the (negative) heating rate of devices in the \texttt{on} state ($v_{\mathrm{on}}(T)=-\alpha(T-\Ton)$) and the heating rate of devices in the \texttt{off} state ($v_{\mathrm{off}}(T)=\alpha(\Toff - T)$). A convenient choice is the net heating rate 
\begin{equation}
v(T,t)=\alpha \beta(t) (T - \Tmax)
\end{equation}
 that is parameterised by a single \emph{control parameter} $\beta(t)$. The effect of this heating rate profile is a temperature distribution that contracts to the pivot temperature $\Tmax$ in order to provide energy to the grid - and reverses this process to recover the energy supplied. In \cite{Trovato2015} it was coupled to a `mirrored' controller that is capable of absorbing energy from the grid by contracting to the pivot temperature $\Tmin$. The controller switches between \emph{energy absorption} and \emph{energy provision} modes whenever $\overline{T}(t)$ crosses $\overline{T}_0$ (when $z(t)$ crosses 0). 

A generalised formulation covering both regimes is obtained by defining a heating rate of the form $v(T,t;R) = \alpha \beta(t; R) (T - R(t))$, where $R \in \{\Tmin, \Tmax \}$ is a reference temperature, which acts as a pivot temperature for the controller, with the property $v(R,t;R)=0$. The reference temperature is defined as follows:
\begin{equation}
R(t)= \begin{cases}
\Tmax, & \text{if } \overline{T}(t) \ge \overline{T}_0 \\
\Tmin, & \text{if } \overline{T}(t) < \overline{T}_0 
\end{cases} 
\end{equation}

\subsection{Control parameter}
The control parameter $\beta(t;R)$ is determined by the desired reference power $\Pi(t)$ according to \cite[Eq.~(36)]{Tindemans2015}:
\begin{align}
\beta(t;R) &=\frac{ \Pi(t)(\Toff - \overline{T}_0)  - (\Toff - \overline{T}(t) ) }{R(t) - \overline{T}(t)} \nonumber \\
&= \frac{(\Pi(t) - 1) - z(t)}{z(t) - \zeta(R(t))}  \label{eq:betaOldDefinition}
\end{align}
where
\begin{equation}
\zeta(R) = \frac{\overline{T}_0 - R}{\Toff - \overline{T}_0}. \label{eq:zetadefinition}
\end{equation}
The denominator in the definition of $\beta$ reflects, in dimensionless form, the energy limits of the TCL aggregate. Note also that $\beta$ switches sign depending on the value of $R(t)$.

\subsection{Distribution scaling}

The controller has the effect of scaling the steady state temperature distribution $f_0(T)$ around the pivot temperature $R(t)$, such that the distribution does not exceed the temperature bounds $\Tmin$ and $\Tmax$ \cite{Tindemans2015}. The extent of this scaling at time $t_i$ is compactly represented by the scale parameter
\begin{align}
s(t) &= \frac{R(t) - \overline{T}(t)}{R(t) - \overline{T}_0} \nonumber \\
&= 1- z(t)/\zeta(R(t)) . \label{eq:sdefinition}
\end{align}

\subsection{Discretisation} \label{sec:discretisation}

We now consider the restriction of the continuous time controller to the set of discrete times $t_i$. We replace the coordinate $z(t)$ by its discretisation $z_i = z(t_i)$, which is computed from \eqref{eq:avgtempdefinition} as 
\begin{equation}
z_i = \sum_{j=0}^{i} (\Pi_j - 1) \left( \e{-\alpha (t_i-t_j)} - \e{-\alpha (t_i - t_{j-1})} \right). 
\end{equation}
Updates to $z_i$ are efficiently implemented using $z_0=0$ (for a steady state initialisation) and the recursive relation
\begin{equation} \label{eq:zupdate}
z_i = z_{i-1} \e{-\alpha \Delta t_i } + (\Pi_i - 1) (1- \e{-\alpha \Delta t_i }).
\end{equation}

The discretised controller switches modes only at instants $t_i$, so $R(t)$ is approximated by the delayed function
\begin{equation}
\hat{R}(t) = R_i, \quad \text{for } t\in (t_{i-1}, t_i]. 
\end{equation}
with
\begin{equation} \label{eq:Rip1definition}
R_{i+1} = \begin{cases}
\Tmax, & \text{if } z_i \le 0, \\
\Tmin, & \text{if } z_i >0.
\end{cases}
\end{equation}
Because our analysis focuses on the control time $t_i$, where $\hat{R}(t)$ and $\Pi(t)$ are potentially discontinuous, we introduce $\pm$-notation for the left and right limits at $t_i$: 
\begin{subequations}
\begin{align}
R_i^- &= \lim_{\varepsilon \downarrow 0}  \hat{R}(t-\varepsilon) = R_i, \\
R_i^+ &= \lim_{\varepsilon \downarrow 0}  \hat{R}(t+\varepsilon) = R_{i+1}.
\end{align}
\end{subequations}
Similar definitions using left and right limits naturally apply to $\zeta(R)$, $s(t)$ and $\beta(t;R)$:
\begin{align} 
\zeta^{\pm}_{i} & =\frac{\overline{T}_0 - R^{\pm}_{i}}{\Toff - \overline{T}_0}, &
 s^{\pm}_{i} &= 1 - z_i /  \zeta_i^{\pm}, \label{eq:zetasdiscrete} \\
\beta^{-}_{i} &=\frac{(\Pi_i - 1) - z_i}{z_i - \zeta_i^-}, &
\beta^{+}_{i} & = \frac{(\Pi_{i+1} - 1) - z_i}{z_i - \zeta_{i}^+}. \label{eq:betasdiscrete}
\end{align}

\subsection{Energy constraints} \label{sec:energyconstraints}

The ability of the aggregate appliances to sustain a low or high power level is determined by operating temperature bounds of the appliance, applied to the distribution-averaged temperature: $\overline{T}(t) \in (\Tmin, \Tmax)$ (no feasible solutions for the distribution $f_z(T)$ exists outside of this domain). However, operation near the limits is infeasible in practice, because appliances would need to switch at very high rates to operate in a narrow temperature range. Therefore, we shall use a restricted range of operating temperatures that is scaled with a fraction $w<1$ around the steady state operating temperature $\overline{T}_0$:
\begin{equation}
(1-w)\overline{T}_0 + w \Tmin \le \overline{T}(t) \le (1-w)\overline{T}_0 + w \Tmax. 
\end{equation}
Rewriting this in terms of $z(t)$ and $\zeta(\cdot)$, we get
\begin{equation}
w \zeta(\Tmax)  \le z(t) \le w \zeta(\Tmin). 
\end{equation}
Small excursions out of this temperature band will be permitted, but if this happens, the requested power level $\Pi_{i+1}$ will be restricted to not exacerbate the excursion, using the relation~\eqref{eq:zupdate}. This leads to the update rule for $\Pi_{i+1}$:
\begin{equation}
\Pi_{i+1} := \begin{cases}
\max (\Pi_{i+1}, 1+ w \zeta(\Tmax)), & \text{if } z_i \le w \zeta(\Tmax)\\
\min (\Pi_{i+1}, 1 + w \zeta(\Tmin) ), & \text{if } z_i \ge w \zeta(\Tmin)\\
\Pi_{i+1}, & \text{otherwise} 
\end{cases}
\end{equation}
Note that $\beta^{+}_i$ will need to be (re-)computed to ensure consistency with an adjusted power level.

\subsection{Power constraints} \label{sec:powerconstraints}

In addition to energy constraints related to the distribution-averaged temperature, the controller is subject to instantaneous power constraints that result from the maximum rate of change of the distribution. Following the procedure in \cite{Tindemans2015}, these limits can be derived by requiring that the desired \emph{average} rate of temperature change for devices with temperature $T_i$ remains within the physical limits imposed by the \texttt{on} and \texttt{off} states, i.e. $v_{\textrm{on}}(T_i) \le v(T_i, t_i) \le v_{\textrm{off}}(T_i)$ for all $T_i$ and $t_i$. Expanding these for the pivoting controller at time $t_i+\varepsilon$ yields
\begin{equation} \label{eq:velocityrange}
 - \alpha (T_i - \Ton) \le \alpha \beta^+_i (T_i - R_i^+) \le - \alpha (T_i - \Toff).
\end{equation}

Since this equality is linear in $T_i$ and must hold for all devices, it suffices to verify this constraint at the most extreme attainable temperatures. From the linear scaling of the temperature distributions around the pivot temperature $R(t)$ with a factor $s(t)$, it follows that the permitted temperature interval $[\Tlow(t), \Thigh(t)]$ is defined by
\begin{subequations}
\begin{align}
\Tlow(t) = R(t) - (R(t)  - \Tmin)s(t), \\
\Thigh(t) = R(t) - (R(t) - \Tmax) s(t). 
\end{align}
\end{subequations}
Depending on the controller mode, one of these limits is equal to the pivoting temperature $R(t)$. Since we have $v(R(t),t)=0$ by definition, \eqref{eq:velocityrange} is trivially satisfied at this limit. We can therefore focus on the opposite limit, which can be written as
\begin{equation}
T_{\mathrm{limit}}(t)= R(t) + \left[\Tmin + \Tmax - 2 R(t) \right]s(t).
\end{equation}
Setting $T_i=T_{\mathrm{limit}}(t_i+\varepsilon)$ in \eqref{eq:velocityrange} and reordering of the result yields the following power constraints on $\Pi_{i+1}$:
\begin{multline}\label{eq:instantaneousPowerLimits}
1 + \zeta^+_i \left(\frac{\Tmin+\Tmax-\Toff-R^+_i}{\Tmin + \Tmax - 2 R^+_i}\right) \le \Pi_{i+1} \le \\ 1 + \zeta^+_i \left(\frac{\Tmin+\Tmax-\Ton-R^+_i}{\Tmin + \Tmax - 2 R^+_i}\right).
\end{multline}
This result unifies the separately calculated limits for the energy-provision ($z_i < 0$) and energy-absorption ($z_i >0$) regimes in \cite{Tindemans2015, Trovato2015}.

\section{Controller: device switching} \label{sec:switching}
The desired evolution of the temperature distribution can be used to compute the necessary control actions of individual appliances. This section identifies such control actions using the three types of switching events identified in Section~\ref{sec:drcontrol}. These are computed as a function of the time of evaluation $t_i$, the compressor state $c_i \in \{0,1 \}$ during the preceding interval $(t_{i-1}, t_i]$, and the current device temperature $T_i$ (assumed to be measured in the appliance at time $t_i$). 

\subsection{Forced switching}

TCLs are forced to switch \texttt{on} or \texttt{off} when their temperatures exceed the permitted interval $[\Tlow(t), \Thigh(t)]$. From the linear scaling of the temperature distributions around the pivot temperature $R(t)$ with a factor $s(t)$, it follows that 
\begin{subequations}
\begin{align}
\Tlow(t) = R(t) - (R(t)  - \Tmin)s(t), \\
\Thigh(t) = R(t) - (R(t) - \Tmax) s(t). 
\end{align}
\end{subequations}
At $t_i$, the refrigerator must act if these bounds are violated at the start of the next time interval:
\begin{subequations}
\begin{align}
\left[T_i \le R_i^+ - (R_i^+  - \Tmin) s^+_i \right] & \Rightarrow c_{i+1} := 0, \\
\left[T_i \ge R_i^+ - (R_i^+  - \Tmax) s^+_i \right] & \Rightarrow c_{i+1} := 1. 
\end{align}
\end{subequations}

\subsection{Continuous-time switching}

We now consider the continuous-time stochastic switching rates from \texttt{on} to \texttt{off} states ($r^{1\rightarrow 0}(t)$) and vice versa ($r^{0 \rightarrow 1}(t)$), required to maintain the desired shape of the temperature distribution. The switching rates for the energy provision mode are defined in \cite[Eqs.~(48)-(52)]{Tindemans2015}. Here, we generalise these expressions to cover both energy provision and absorption modes ($R(t) \in \{  \Tmin, \Tmax\}$) and simplify them using the $z$-coordinate transformation. Finally, we specialise the expressions for trapezoidal integration with piecewise constant power references.

The derivative of $\beta$ can be simplified by substitution using \eqref{eq:zdefinition} and \eqref{eq:zetadefinition}, resulting in:
\begin{align}
\frac{\mathrm{d}\beta(t;R)}{\mathrm{d} t} &=   \frac{1}{z(t) - \zeta(t)}\frac{\mathrm{d}\Pi(t)}{\mathrm{d}t} + \alpha \beta(t;R) \frac{1 + \zeta(t) - \Pi(t)}{z(t) - \zeta(t)} \nonumber \\
&= \frac{1}{z(t) - \zeta(t)}\frac{\mathrm{d}\Pi(t)}{\mathrm{d}t} - \alpha \beta(t;R) (1+\beta(t;R)) 
\end{align}
This substitution can be used in \cite[Eq.~(51)]{Tindemans2015} (adjusted for general $R$). Further simplification follows from setting $\mathrm{d}\Pi(t)/ \mathrm{d}t = 0$ (because we consider piecewise constant sections between $t_i$). We compute the intermediate quantity $\Xi(t)$, using the identity found in \cite[Eq.~(38)]{Tindemans2015}, again taking left and right limits due to discontinuity at $t_i$.
\begin{subequations}\label{eq:xidefinition}
\begin{align} 
\Xi^{\pm}_i =& \lim_{\varepsilon \downarrow 0}\Xi(T_i,t_i\pm \varepsilon) \nonumber \\
=& \alpha^2 \left(\frac{P^{\pm}_i + Q^{\pm}_i}{P^{\pm}_i Q^{\pm}_i} \right) (X^{\pm}_i Y^{\pm}_i)  -  \alpha^2 (1+\beta^{\pm}_i) (X^{\pm}_i + Y^{\pm}_i) 
\end{align}
with
\begin{align}
P^{\pm}_i &= (R^{\pm}_i - \Toff) s^{\pm}_i + (T_i - R^{\pm}_i)\\
Q^{\pm}_i &= (R^{\pm}_i - \Ton) s^{\pm}_i + (T_i - R^{\pm}_i)\\
X^{\pm}_i &= (T_i - \Toff) + (T_i - R^{\pm}_i) \beta^{\pm}_i\\
Y^{\pm}_i &= (T_i-\Ton) + (T_i - R^{\pm}_i) \beta^{\pm}_i 
\end{align}
\end{subequations}
The stochastic transition rates at the left and right limits to $t_i$ are computed from $\Xi^{\pm}_i$ using \cite[Eq.~(49)]{Tindemans2015} (adjusted for general $R$), resulting in
\begin{subequations} \label{eq:rateequations}
\begin{align}
r^{1\rightarrow 0}_{i,\pm} &= \max\left(0,- \frac{\Xi^{\pm}_i}{\alpha X^{\pm}_i} \right) \\
r^{0 \rightarrow 1}_{i,\pm} &= \max\left(0,- \frac{ \Xi^{\pm}_i}{\alpha Y^{\pm}_i} \right) 
\end{align}
\end{subequations}

Midpoint integration between adjacent time instants $t_{i-1}$ and $t_i$ is used to determine the resulting switching probabilities, where switching is implemented at $t=t_i$:
\begin{subequations} \label{eq:stochastic}
\begin{align}
\text{Pr}^{1  \rightarrow 0}_{\text{cont},i} &= \frac12 \Delta t_i (r^{1  \rightarrow 0}_{i-1,+} + r^{1 \rightarrow 0}_{i,-})  \\ 
\text{Pr}^{0 \rightarrow 1}_{\text{cont},i} &= \frac12 \Delta t_i (r^{0 \rightarrow 1}_{i-1,+} + r^{0 \rightarrow 1}_{i,-}) 
\end{align}
\end{subequations}
Note that the rates at both `inner' edges of the interval $\Delta t_i$ are used: the `$+$' side at $t_{i-1}$ and the `$-$' side at $t_i$.

\subsection{Instantaneous switching}

Finally, consider the instantaneous stochastic switching at time $t_i$ due to mode changes (energy absorption, energy delivery) or changes in $\Pi(t)$. This results in a discontinuous change in the net heating rate $v(T,t)$, which can only be achieved by a fraction of devices switching on or off at $t_i$. Following \cite{Trovato2015}, we compute the probability of switching from the \texttt{on} to \texttt{off} state at time $t_i$, for a refrigerator that is currently \texttt{on}, as
\begin{subequations} \label{eq:instantaneous}
\begin{equation}
\text{Pr}^{1 \rightarrow 0}_{\text{inst},i}=\text{max}\left(0, 1- \frac{X^{+}_i}{X^{-}_i } \right) \label{eq:instantjumpprob1}
\end{equation}
Note that the switching probability includes both a contribution from the discrete change in power level at $t_i$ as well as a possible mode transition in the previous interval that is implemented at $t_i$. The switching probability for fridges in the \texttt{off} state, $c_i=0$, is defined analogously as
\begin{equation}
\text{Pr}^{0 \rightarrow 1}_{\text{inst},i}=\text{max}\left(0, 1- \frac{Y^{+}_i }{Y^{-}_i } \right) \label{eq:instantjumpprob2}
\end{equation}
\end{subequations}

\subsection{Combined stochastic switching}
Formally, the continuous-time \eqref{eq:stochastic} and instantaneous \eqref{eq:instantaneous} switching probabilities should be evaluated in sequence, because the former occurs during the interval $(t_{i-1},t_i]$ and the latter at time $t_i$. This would account for the possibility that an appliance switches off and on again within a single interval, or vice versa. Here, we assume that the switching probability associated with the continuous-time process is small to allow us to evaluate both probabilities in a single step. 
\begin{subequations} \label{eq:finalprobabilities}
\begin{align} 
\text{Pr}^{1 \rightarrow 0}_i &=  \text{Pr}^{1  \rightarrow 0}_{\text{cont},i} +  \text{Pr}^{1 \rightarrow 0}_{\text{inst},i} \\
\text{Pr}^{0 \rightarrow 1}_i & = \text{Pr}^{0 \rightarrow 1}_{\text{cont},i} + \text{Pr}^{0 \rightarrow 1}_{\text{inst},i}
\end{align}
\end{subequations}

\section{Algorithm and implementation} \label{sec:algorithm}

The discrete time algorithm for updating the compressor state derived in sections~\ref{sec:distribution} and \ref{sec:switching} is summarised in pseudocode in Algorithm~\ref{alg:update}. The algorithm was implemented in Python 3.7.7 using the \texttt{numba} package to benefit from just-in-time compilation for considerable speedups. The source code for the controller and all results in this paper has been released under the MIT license \cite{Tindemans2020a}; an interactive compute capsule that reproduces all figures and numerical results is available at \cite{Tindemans2020}. 

\begin{figure}[!t]
\centering
\includegraphics{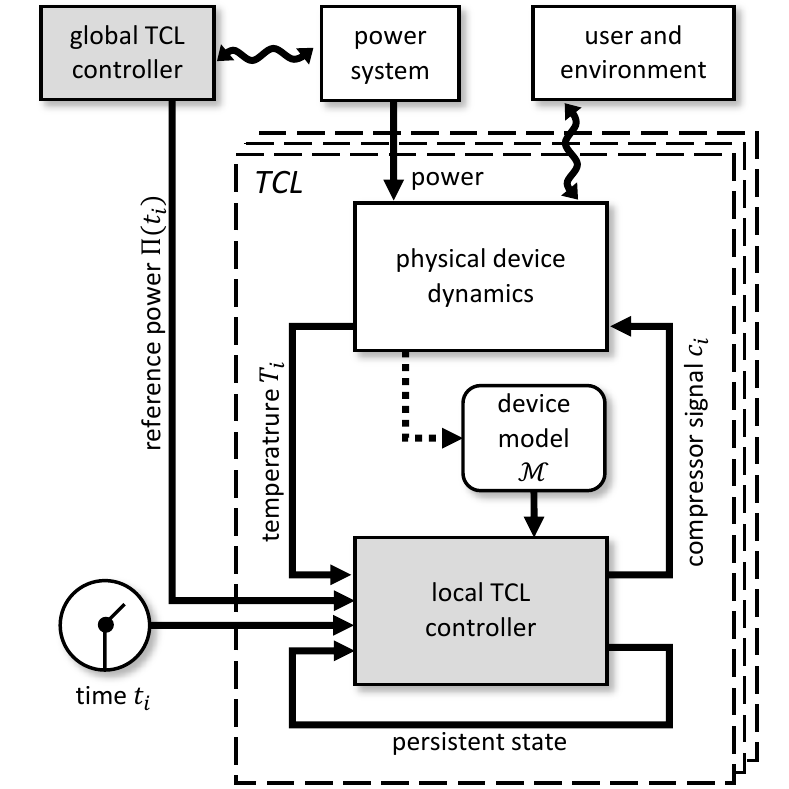}
\caption{Schematic depiction of the local and global control blocks.}
\label{fig:schematic}
\end{figure}

Figure~\ref{fig:schematic} provides a schematic description of the relation between the control algorithm (\emph{local TCL controller} block) and the other components of the overall system. The behaviour of each TCL is determined by the interaction between its physical device characteristics, its environment (e.g. ambient temperature and door openings) and the compressor signal $c_i$. The global TCL controller -- which could be a transmission system operator or a demand response service provided -- monitors the condition of the power system or its associated markets and sends the dispatch signal $\Pi(t_i)$ to all TCLs. 

The TCL controller makes use of this signal, in combination with the current time $t_i$, the measured cabinet temperature $T_i$. It further uses a persistent controller state, and a representation of the device model $\mathcal{M}$. Ideally, this model is a good representation of the physical device characteristics, including its environment and usage. In practice, there will be parameter and model errors to contend with, leading to unavoidable control errors. In advanced implementations, the model can be refined on the basis of observed temperature measurements. 

Due to the conceptual separation between the state update control algorithm and the physical evolution of the system, the control algorithm is equally suitable for embedded applications as for simulations. When embedded in a real system, all other blocks except for the global TCL controller are physical. In a simulation environment, the other blocks are all provided by the modeller. 

\begin{algorithm}
\caption{State update algorithm}\label{alg:update}
\begin{algorithmic}[5]
\Function{update\_compressor\_state}{$\Pi_{i+1}, T_i, t_i$}
	\State \emph{\# load previously computed information}
	\State load appliance model $\mathcal{M}$ and operating range $w$
	\State load $c_i, \Pi_i, z_{i-1}, t_{i-1}, r^{1\rightarrow 0}_{i-1,+} , r^{0\rightarrow 1}_{i-1,+} $ 
	\State
	\State \emph{\# update population parameters}
  	\State compute $z_i$ using \eqref{eq:zupdate} 
	\State compute $R^{\pm}_{i}, \zeta^{\pm}_{i}, s^{\pm}_i$ using \eqref{eq:Rip1definition}-\eqref{eq:zetasdiscrete}
	\State
	\State \emph{\# implement power and energy limits}
	\If{$z_i \le w \zeta(\Tmax)$} \Comment{lower energy limit violated}
		\State $\Pi_{i+1} \gets \max (\Pi_{i+1}, 1+ w \zeta(\Tmax))$
	\EndIf
	\If{$z_i \ge w \zeta(\Tmin)$} \Comment{upper energy limit violated}
		\State $\Pi_{i+1} \gets \min (\Pi_{i+1}, 1 + w \zeta(\Tmin))$
	\EndIf
	\State clip $\Pi_{i+1}$ to power limits in \eqref{eq:instantaneousPowerLimits} 
	\State compute $\beta^{\pm}_i$ using \eqref{eq:betasdiscrete}
	\State
	\State \emph{\# determine distribution and switching variables}
	\State compute $r^{1\rightarrow 0}_{i,\pm}, r^{0\rightarrow 1}_{i,\pm}$  using \eqref{eq:xidefinition}-\eqref{eq:rateequations}
	\State compute $\text{Pr}^{1 \rightarrow 0}_i, \text{Pr}^{0 \rightarrow 1}_i$ using \eqref{eq:stochastic}-\eqref{eq:finalprobabilities}
	\State
	\State \emph{\# implement compressor switching}
	\If{$c_i = 1$} \Comment{currently \texttt{on}}
		\If{$T_i \le R_i^+ - (R_i^+ - \Tmin)s^+_i$} $c_{i+1} \gets 0$ 
		\Else
			\State $u \gets \textrm{uniform random}\in [0,1]$
			\If{$u \le \text{Pr}^{1 \rightarrow 0}_i$} $c_{i+1} \gets 0$
			\Else ~$c_{i+1} \gets 1$ \Comment{remain \texttt{on}}
			\EndIf
		\EndIf
	\Else \Comment{currently \texttt{off}}
		\If{$T_i \ge R_i^+ - (R_i^+ - \Tmax)s^+_i$} $c_{i+1} \gets 1$
		\Else
			\State $u \gets \textrm{uniform random}\in [0,1]$
			\If{$u \le \text{Pr}^{0 \rightarrow 1}_i$} $c_{i+1} \gets 1$
			\Else ~$c_{i+1} \gets 0$ \Comment{remain \texttt{off}}
			\EndIf
		\EndIf
	\EndIf
   \State \textbf{return} $c_{i+1}$ \Comment{updated compressor state}
\EndFunction
\end{algorithmic}
\end{algorithm}

\section{Results}\label{sec:results}

For simulations, thermal model parameters were taken from \cite[domestic refrigerator class]{Trovato2015}: $\alpha=1/7200 s^{-1}$; $\Tmax=7\degree C$; $\Tmin=2 \degree C$; $\Ton=-44\degree C$; $\Toff=20 \degree C$. Heterogeneous appliances were generated from these parameters by individually multiplying their values with a random factor that was uniformly distributed between $0.8$ and $1.2$. The temperature evolution of each appliance was computed by integrating \eqref{eq:ode-physics} using the Euler method; the integration time steps equaled those of the controller, unless otherwise noted. All appliances had a maximum power consumption $P^a_{\mathrm{on}}=70W$ and operating range $w=0.9$ (not binding for the parameters used). Each appliance was randomly initialised as follows. The compressor was set to the \texttt{on} state with a probability equal to the steady state duty cycle $P^a_0/P^a_{\mathrm{on}}$ and the temperature was initialised in the range $[\Tmin, \Tmax]$ according to the steady state probability distributions $f_0(T | c^a_0=1) \propto 1/(T-\Ton)$ and $f_0(T | c^a_0=0) \propto 1/(\Toff - T)$. 

\begin{figure}[!t]
\centering
\includegraphics{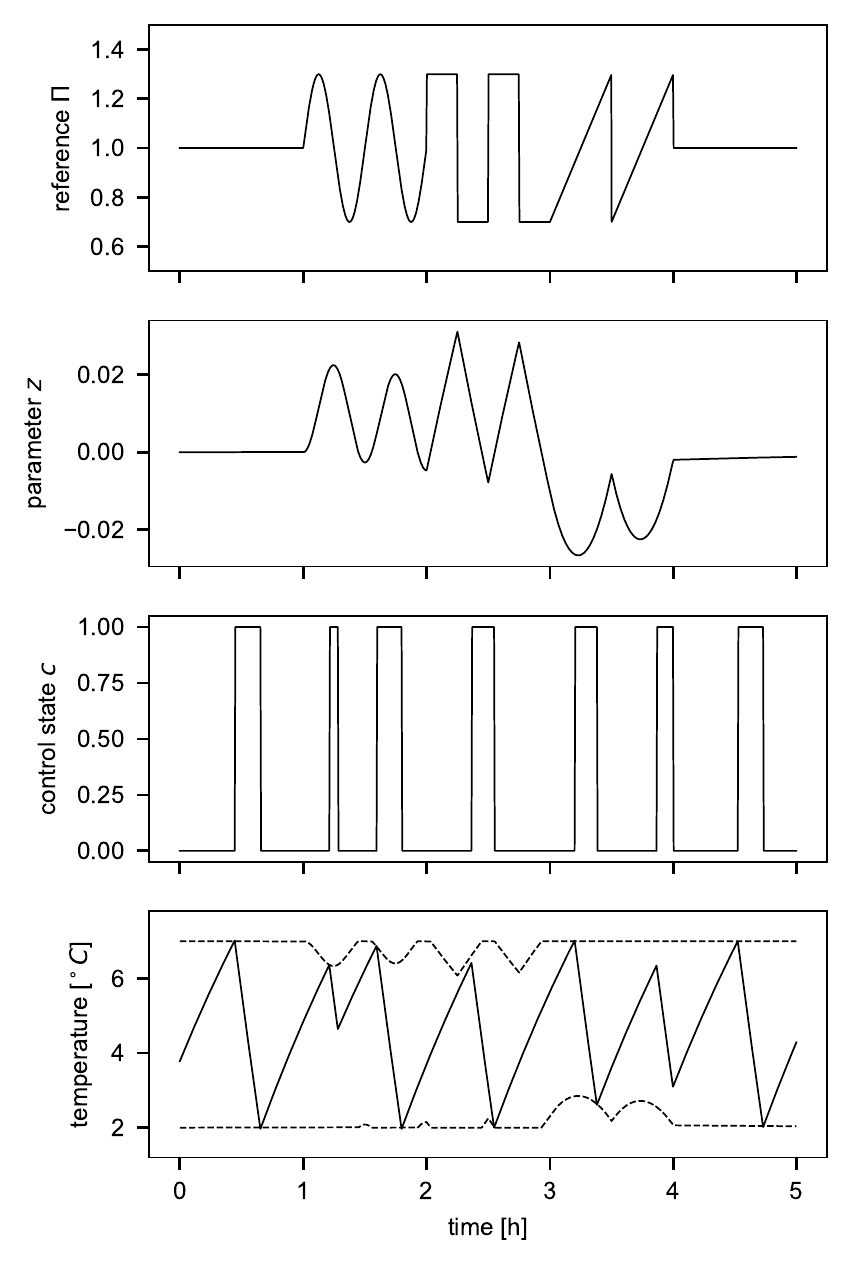}
\caption{Reference signal (top) and corresponding population parameter (second). The bottom two panels show the state evolution of a random single applicance (compressor state and temperature). 10s time steps were used for the simulation.}
\label{fig:single}
\end{figure}

\subsection{Single device analysis}

Figure~\ref{fig:single} shows a reference signal (top), with a length of 5 hours, that demonstrates a variety of features. The second panel shows the resulting value of the population parameter $z$, indicating that the controller is predominantly in the energy-absorption mode ($z \ge 0$) during the first 3 hours, before switching to the energy-provision mode ($z\le 0$) for the remainder of the time. 

The bottom two panels show the compressor state $c_i$ and temperature $T_i$, respectively, of a single appliance that tracks the reference signal. The control state of a single appliance does not have a very apparent relation with the reference signal. The bottom panel indicates the device temperature $T_i$ (solid line) alongside the lower and upper bound temperatures $\Tlow(t)$ and $\Thigh(t)$ (dashed lines). This illustrates the ability of the controller to strictly respect the temperature bounds. It also demonstrates that \texttt{on}/\texttt{off} switching events occur both due to the device reaching its temperature bounds (deterministic) and due to stochastic switching events at intermediate temperatures.

\begin{figure}[!t]
\centering
\includegraphics{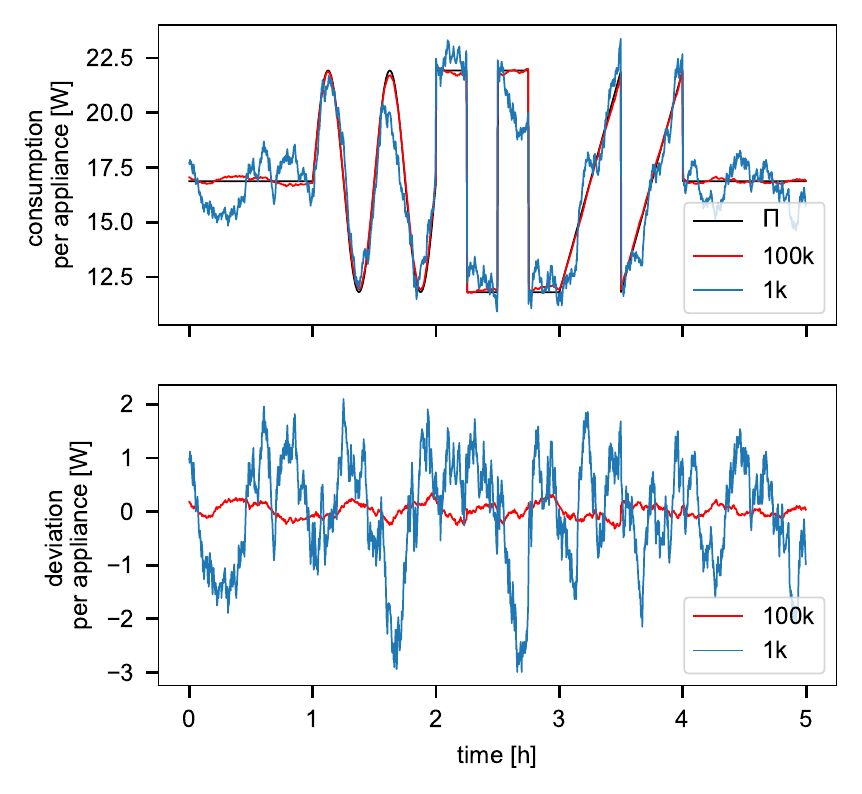}
\caption{Response of a heterogeneous aggregate of appliances (top) and deviation from the reference (bottom). 10s time steps were used for the simulation.}
\label{fig:multi}
\end{figure}

\subsection{Population size}
Next, heterogeneous populations of $1{,}000$ and $100{,}000$ appliances were simulated, tracking the same reference signal. Figure~\ref{fig:multi} illustrates the convergence of the aggregate response to the reference signal as the number of independent appliances increases. The top panel shows absolute power consumption per appliance; the bottom panel the deviation from the reference in Watt per appliance. 

To further analyse the convergence as a function of device counts, the signal was repeated five consecutive times to obtain a simulated 25-hour run. The per-appliance power deviation for these runs amounted to $0.864~W$ and $0.116~W$, respectively, for populations of $1{,}000$ and $100{,}000$ appliances. Although this reduction in standard deviation is substantial ($7.4\times$), it nevertheless falls short of the $10\times$ reduction predicted by \eqref{eq:sumconvergence}. This discrepancy can largely be attributed to the 10-second step size used for the simulations: devices that exceed temperature bounds or are affected by a stochastic switching event must wait until the next 10-second barrier to implement this event. Simulating $100{,}000$ appliances with 1-second time steps (and a reference signal with stepwise increments at 10-second intervals, for a fair comparison) resulted in a standard deviation of only $0.098~W$ per appliance. This number can be compared to the total signal magnitude of approximately $10~W$ per appliance.

\begin{figure}[!t]
\centering
\includegraphics{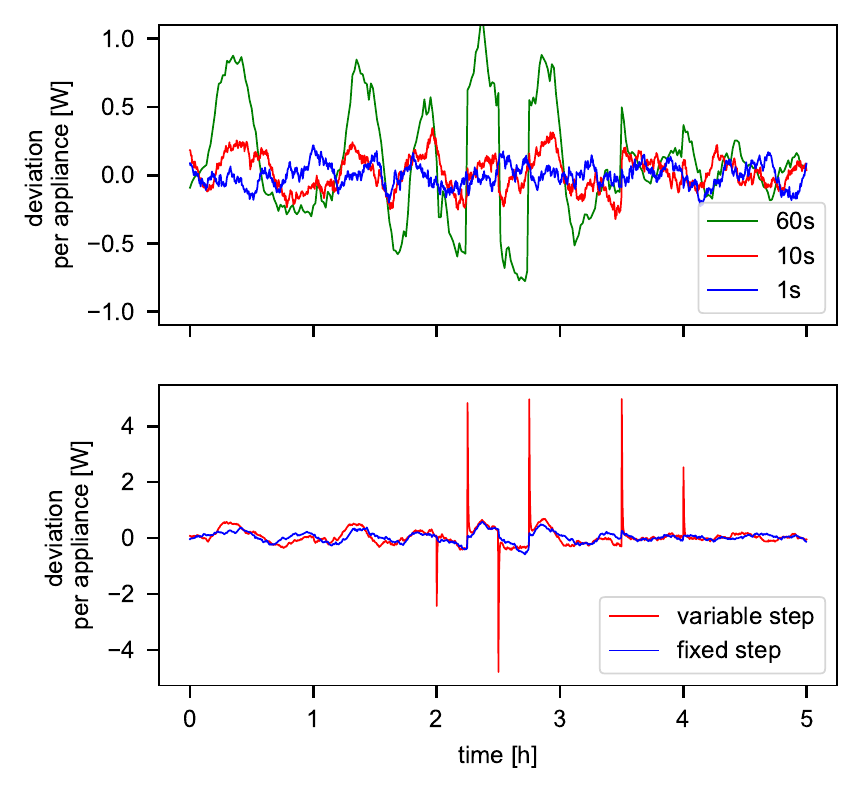}
\caption{Deviation between the reference signal from Fig.~\ref{fig:multi} and response of $100{,}000$ appliances for various time steps. The top panel compares fixed time steps of various sizes. The bottom panel compares a fixed 20s time step with a 10s time step where half of controller invocations is dropped at random.}
\label{fig:timestep}
\end{figure}

\subsection{Simulation step size} \label{sec:stepsize}

The impact of the time step size is further investigated in Figure~\ref{fig:timestep}. The top panel compares the reference tracking accuracy for time steps of 1, 10 and 60 seconds. Whereas 60s time steps result in significant `cross-talk' from the reference signal on the deviation signal, this is much reduced for 10s time steps and nearly absent (at least visually) for 1s time steps. This suggests that 10s time steps are sufficient for all but the most critical applications (for this population size of 100,000 appliances).

The bottom panel compares the aggregate tracking accuracy obtained using fixed time steps of 20s and variable time steps with an \emph{average} duration of 20s. Variable time steps were generated by running a simulation with 10s time steps and skipping each time step with 50\% probability -- independently for each time step and device. Fig.~\ref{fig:timestep} shows that the tracking performance is similar, with the exception of discontinuous transitions in the reference power, where some devices with variable time steps are unable to instantly track the desired change. However, the tracking fully recovers after a short delay. Moreover, the fixed time step controller \emph{is} able to track all discontinuities in the reference signal instantaneously.

\begin{figure}[!t]
\centering
\includegraphics{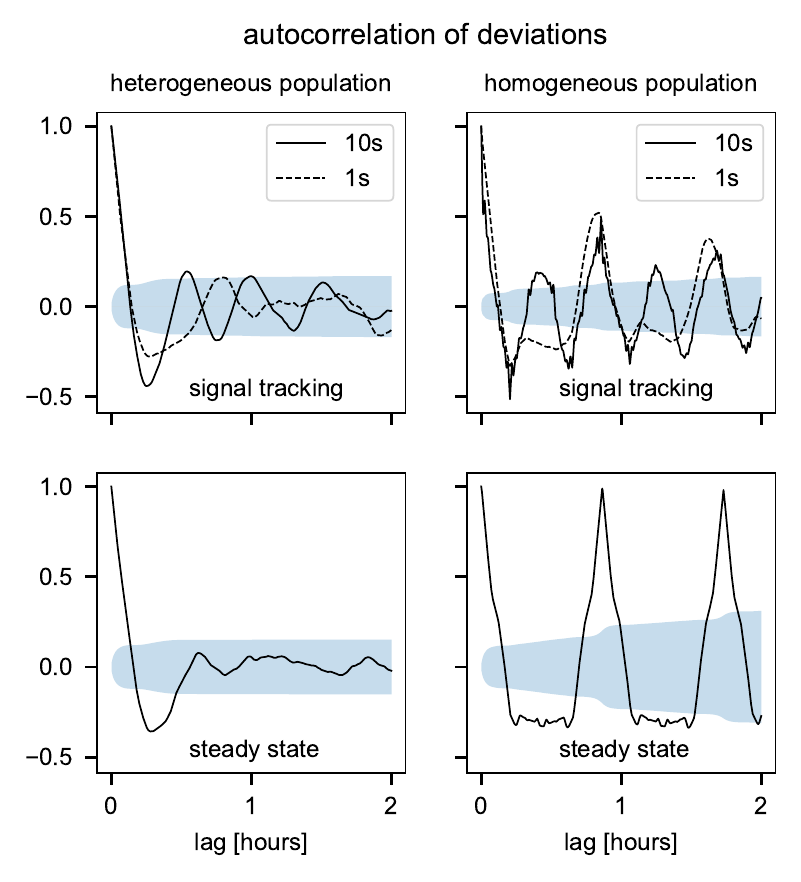}
\caption{Autocorrelation results obtained with populations of 100,000 appliances and 25-hour simulation runs. Top panels are show results for signal tracking simulations; bottom panels for steady state simulations. Solid lines indicate results from 10-second time steps and dashed lines using 1-second time steps. Values within shaded areas are non-significant at the 95\% confidence level.}
\label{fig:autocorrelation}
\end{figure}

\subsection{Temporal correlations of tracking errors}
In this section, we look closer at the temporal signature of the tracking errors using their autocorrelation, shown in Fig.~\ref{fig:autocorrelation}. The panels on the right show that, for homogeneous populations of TCLs, long-range correlations are present in the deviation from theoretical mean, even resulting in \emph{exact} repetition in the steady state. However, when the population tracks a signal (top right), this long-range autocorrelation is reduced, because stochastic switching effectively randomises the appliance states. 

For heterogeneous populations, the autocorrelation signatures are much reduced, especially in the steady state. However, residual long-range correlations were still observed for the signal tracking case (top left; solid line), suggesting that they were induced by the control signal. 

The solid lines are computed on the basis of 10s time steps, which were shown in Section~\ref{sec:stepsize} to be marginally sufficient for good tracking performance. Additional results using 1s time steps (dashed lines in both top panels) show that this indeed reduces the long-term autocorrelations for the heterogeneous case, but not for the homogeneous population case. Moreover, the periodicity of both signals changes from that of the control signal (approx.~30 minutes) to that of the average cycle length of the appliances (approx.~1 hour), in line with the steady state results. This supports the hypothesis that a reduction in time step to 1s is sufficient to eliminate the impact of the control signal for critical applications.

\subsection{Model error}

The proposed control algorithm relies on a thermal model of the appliance to compute the appropriate switching parameters. In the previous sections, we have assumed that this model was known to the controller, but this is arguably a strong assumption that does not hold in practice \cite{Kara2015}. The sensitivity to model errors is investigated in this section. 

For this purpose, we distinguish the \emph{physical model} that dictates the appliance physics according to \eqref{eq:ode-physics} and the \emph{control model} that is used in Algorithm~\ref{alg:update}. Three cases were considered:
\begin{itemize}
\item Known models. The control model parameters were identical to those of the physical model. 
\item Common control model. The physical model for each device was randomly drawn as described above, but devices share a nominal control model (with median parameter values). The latter can be thought of as a factory-embedded `archetype model' for the appliance type. 
\item Randomized model errors. Parameters for the physical and control models were drawn independently. Such a scenario might occur if each appliance independently (and imperfectly) attempts to learn its own model.
\end{itemize}
In all cases, the control parameters $\Tmin^a$ and $\Tmax^a$ were taken from the physical model, as they represent the user's cooling preferences, as adapted to the physical environment of the appliance. The physical model parameters were used to generate the initial random temperature of the device, which represents an initial state where a hysteresis controller was used prior to $t=0$. 

\begin{figure}[!t]
\centering
\includegraphics{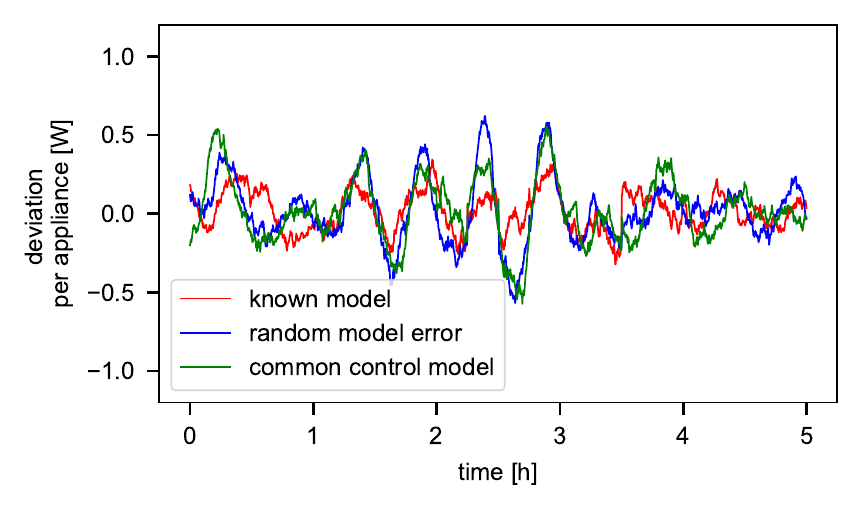}
\caption{Comparison of aggregate tracking error of heterogeneous appliances for different types of model error. Three cases are depicted: known models (red; also shown in Fig.~\ref{fig:multi} and Fig.~\ref{fig:timestep}), randomised model errors (blue), and the use of a common control model (green). Simulations used 100,000 devices and 10$s$ time steps.}
\label{fig:modelerror}
\end{figure}

Figure~\ref{fig:modelerror} compares the average tracking error for all three scenarios, when tracking the signal shown in Fig.~\ref{fig:single} (top) with 100,000 devices. Model error is shown to slightly increase the tracking error, especially when following a discontinuous signal (hours 2-3). However, the magnitude of this error remains well-controlled and the error signal quickly reduces to normal levels when the reference signal reverts to the nominal value $\Pi(t)=1$. We emphasise that model parameters may be off by as much as 50\% in the case of random model error, further illustrating the robustness of the controller design. However, we point out that a significant bias in the parameter estimation errors may lead to a deterioration of performance, as errors are less likely to cancel out. 

\subsection{Door openings}

The model errors analysed in the previous section represent a constant discrepancy between the assumed thermal model and reality. In addition, it is worthwhile to consider  \emph{dynamic} disturbances such as the effect of changing outdoor temperatures \cite{Hreinsson2020} and the impact of frequent door openings \cite{Vrettos2016b}. In the following, we consider only the case of door openings, which induce sudden large changes in the model state (temperature). As such, they can be thought of as a severe stress test of the controller. 

\begin{figure*}[!h]
\centering
\includegraphics{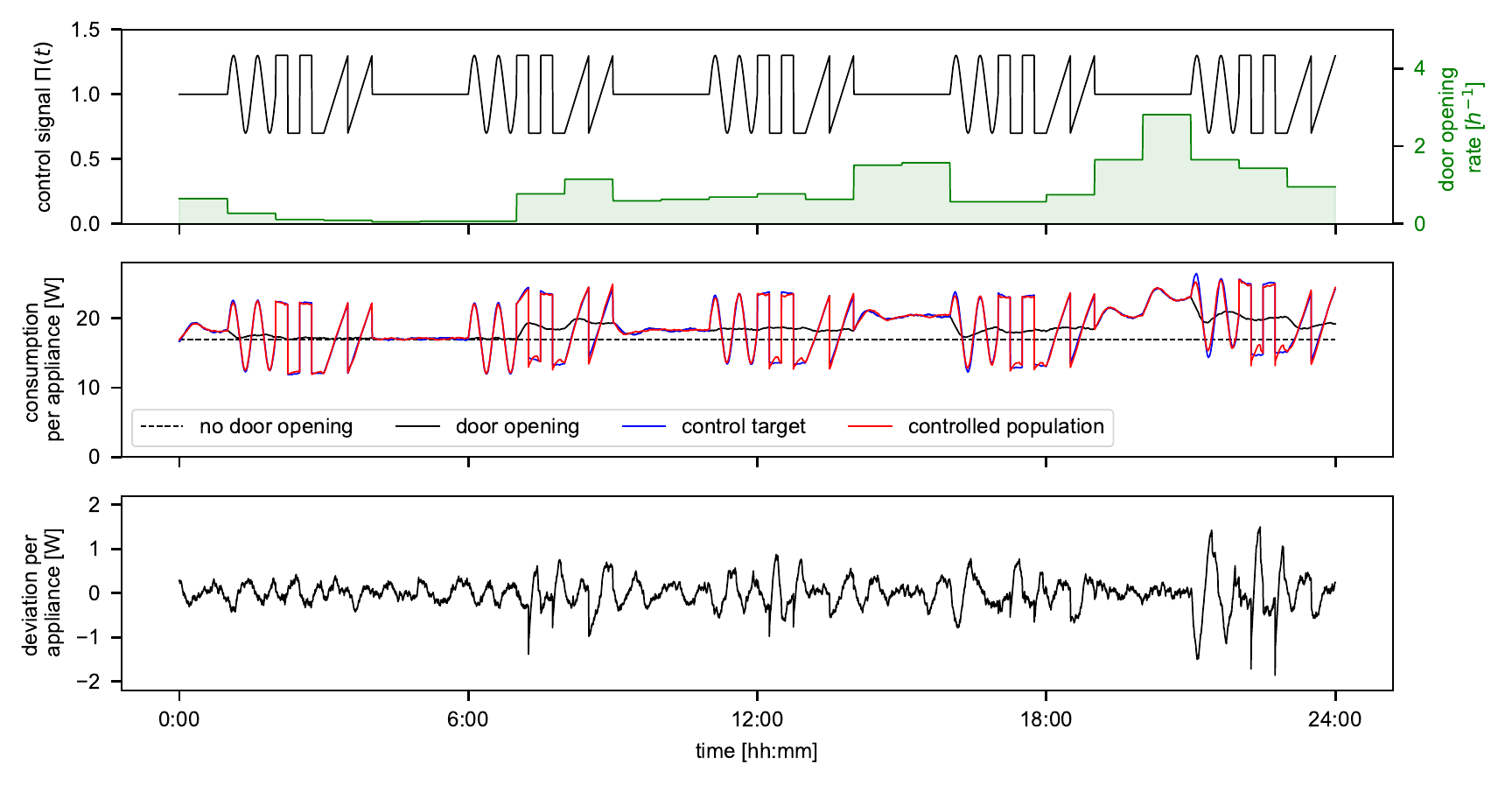}
\caption{Results of an aggregate control experiment in the presence of door openings. Top: 24-hour reference signal (a repeated test pattern) and the door opening rate with hourly changes. Middle: comparison of per-appliance power consumption with different baselines. Bottom: per-appliance tracking error between the controlled population and the control target.}
\label{fig:dooropening}
\end{figure*}

We model door openings to occur independently with a time-varying rate, adapted from \cite[Fig.~15]{Vrettos2016b} and scaled to result in an average of 20 openings per day (see Fig.~\ref{fig:dooropening}, top panel).  Each door opening event lasts 20 seconds and results in a reduction of the thermal conductance $R^a$ by a factor of 25 \cite{Vrettos2016b} while the door is open. For the simulation of temperature dynamics, the time step of Euler integration was reduced by the same factor of 25 to reduce numerical errors, but the time step for the controller remained the same at 10 seconds. 

The results of this simulated experiment with 100,000 devices are shown in Figure~\ref{fig:dooropening}. The middle panel shows the expected power consumption (per appliance) in absence of door openings (dashed black line), alongside the simulated power consumption with door openings (solid black line). The first observation is that the addition of door openings leads to a significant and time-varying increase in average power consumption, by 11\% on average and 44\% at its peak. 

The red line indicates the power consumption of appliances with door openings that simultaneously track the reference signal shown in Fig.~\ref{fig:dooropening} (top). In order to quantify the tracking performance, this should be compared with a target power consumption, for which the reference signal $\Pi(t)$ would be designed. The time-varying door openings result in a time-varying baseline for the power consumption, so the target power consumption can be given by
\begin{equation}
\bar{P}^{\textrm{target}}_{\Pi, \textrm{door}} (t) = \bar{P}_{0,\textrm{door}}(t) + (\Pi(t) - 1) \bar{P}_0, 
\end{equation}
where all powers $\bar{P}_x$ are taken to be averages over the population (i.e. $W$ per device). The second term specifies the modulation of the power consumption with respect to the known steady state power level $\bar{P}_0$. The baseline power consumption $\bar{P}_{0,\textrm{door}}(t)$ is variable, but it can be predicted in the long term. As no analytical expression is available for this baseline, the empirical power consumption (solid black line) for an uncontrolled population is used instead. The resulting target power consumption $\bar{P}^{\textrm{target}}_{\Pi, \textrm{door}} (t)$ is indicated by the blue line in the middle panel of Fig.~\ref{fig:dooropening}. The bottom panel depicts the difference between this reference power consumption and the realised power consumption. It demonstrates that the controller is able to track the signal to within an error of $\pm 2 W$/appliance, despite very large perturbations in the form of door openings occurring at the same time as large swings in reference power consumption.  

\section{Conclusions and future work}
This paper has derived a discrete time TCL controller for decentralised demand response. The results illustrate the ability to accurately track a (potentially discontinuous) reference signal using a large population of heterogeneous appliances. The tracking performance remained high even in the presence of significant model errors and random refrigerator door openings that cause unanticipated spikes in appliance temperatures. 

Moreover, from a computational perspective, it was shown that the time-discretisation procedure used led to acceptable tracking errors for time steps of 10 seconds, and further suppression of correlated errors with a smaller time step of 1 second. Coupled with the low computation complexity of Algorithm~\ref{alg:update}, this permits implementation on embedded hardware with severe computational constraints, or it can be used to achieve efficient off-line simulations. The simulation of 100,000 devices for 5 hours using 10s time steps took only 35 seconds (using an Intel i5-7360U CPU under macOS 10.15.4 and Python 3.7.7). Moreover, the ability to use variable time steps can further alleviate real time constraints.

In addition to these computational benefits, the controller inherits desirable properties from its continuous-time version \cite{Tindemans2015} that enable a wide range of potential applications. The ability to accurately track discontinuous reference signals in a decentralised manner provides a natural fit for ancillary services, including primary and secondary frequency response. The efficacy of various control frequency control services that can be implemented on the back of this control mechanism is investigated in \cite{Trovato2017}. 

Moreover, the exact mapping of the aggregate TCL population onto a `leaky battery' representation \cite{Trovato2015} means that both response and recharging can be scheduled optimally according to aggregator objectives. Trovato \emph{et al.} \cite{Trovato2018} have used this to optimally schedule primary and secondary response in the Great Britain system and calculate the associated benefits. The ability of some TCLs (e.g. commercial freezers) to significantly shift their energy consumption also opens up significant opportunities in energy arbitrage. An investigation of the optimal simultaneous allocation of multiple services (ancillary, arbitrage) for different TCL types is contained in \cite{Trovato2015}. 

The controller can therefore be applied to domestic, commercial and industrial cooling and heating appliances with dynamics that are approximately described by \eqref{eq:ode-physics}, and that can be controlled in sufficient quantities to permit mean-field scheduling of the type \eqref{eq:sumconvergence}. However, further work is required to extend the approach presented here to large refrigerators that have multiple compressors, as they do not strictly satisfy the \texttt{on}/\texttt{off} paradigm used here. 

In future work, further analysis - both theoretical and simulation-based - will be done to investigate the robustness of the proposed control approach, including the effect of outdoor temperatures \cite{Hreinsson2020} and lockout constraints \cite{Vrettos2016b, Totu2017}. This may result in enhancements of the controller that improve its performance in a wide range of scenarios. For example, it is interesting to consider how the controller could be enhanced with a means for an appliance to learn and test its own thermal model. Finally, we highlight that there is a substantial research need to define common test cases and to use these to compare the performance of this and other aggregate TCL control approaches published in recent years. 

\bibliographystyle{IEEEtran}

\end{document}